\documentstyle[
	subeqnarray 
	]{mn}

\newif\ifAMStwofonts

\ifoldfss
  \ifCUPmtlplainloaded \else
    \NewTextAlphabet{textbfit} {cmbxti10} {}
    \NewTextAlphabet{textbfss} {cmssbx10} {}
    \NewMathAlphabet{mathbfit} {cmbxti10} {} 
    \NewMathAlphabet{mathbfss} {cmssbx10} {} 
  \fi
  \ifAMStwofonts
    \ifCUPmtlplainloaded \else
      \NewSymbolFont{upmath} {eurm10}
      \NewSymbolFont{AMSa} {msam10}
      \NewMathSymbol{\upi}     {0}{upmath}{19}
      \NewMathSymbol{\umu}     {0}{upmath}{16}
      \NewMathSymbol{\upartial}{0}{upmath}{40}
      \NewMathSymbol{\leqslant}{3}{AMSa}{36}
      \NewMathSymbol{\geqslant}{3}{AMSa}{3E}

    \fi
  \fi
\fi 

\ifnfssone
  \newmathalphabet{\mathit}
  \addtoversion{normal}{\mathit}{cmr}{m}{it}
  \addtoversion{bold}{\mathit}{cmr}{bx}{it}
  \newmathalphabet{\mathbfit} 
  \addtoversion{normal}{\mathbfit}{cmr}{bx}{it}
  \addtoversion{bold}{\mathbfit}{cmr}{bx}{it}
  \newmathalphabet{\mathbfss} 
  \addtoversion{normal}{\mathbfss}{cmss}{bx}{n}
  \addtoversion{bold}{\mathbfss}{cmss}{bx}{n}
  \ifAMStwofonts
    \ifCUPmtlplainloaded \else
      \UseAMStwoboldmath
      \makeatletter
      \new@mathgroup\upmath@group
      \define@mathgroup\mv@normal\upmath@group{eur}{m}{n}
      \define@mathgroup\mv@bold\upmath@group{eur}{b}{n}
      \edef\UPM{\hexnumber\upmath@group}
      \new@mathgroup\amsa@group
      \define@mathgroup\mv@normal\amsa@group{msa}{m}{n}
      \define@mathgroup\mv@bold\amsa@group{msa}{m}{n}
      \edef\AMSa{\hexnumber\amsa@group}
      \makeatother
      \mathchardef\upi="0\UPM19
      \mathchardef\umu="0\UPM16
      \mathchardef\upartial="0\UPM40
      \mathchardef\leqslant="3\AMSa36
      \mathchardef\geqslant="3\AMSa3E
    \fi
  \fi
\fi 

\ifnfsstwo
  \DeclareMathAlphabet{\mathbfit}{OT1}{cmr}{bx}{it}
  \SetMathAlphabet\mathbfit{bold}{OT1}{cmr}{bx}{it}
  \DeclareMathAlphabet{\mathbfss}{OT1}{cmss}{bx}{n}
  \SetMathAlphabet\mathbfss{bold}{OT1}{cmss}{bx}{n}
  \ifAMStwofonts
    \ifCUPmtlplainloaded \else
      \DeclareSymbolFont{UPM}{U}{eur}{m}{n}
      \SetSymbolFont{UPM}{bold}{U}{eur}{b}{n}
      \DeclareSymbolFont{AMSa}{U}{msa}{m}{n}
      \DeclareMathSymbol{\upi}{0}{UPM}{"19}
      \DeclareMathSymbol{\umu}{0}{UPM}{"16}
      \DeclareMathSymbol{\upartial}{0}{UPM}{"40}
      \DeclareMathSymbol{\leqslant}{3}{AMSa}{"36}
      \DeclareMathSymbol{\geqslant}{3}{AMSa}{"3E}
    \fi
  \fi
\fi 

\ifCUPmtlplainloaded \else
  \ifAMStwofonts \else 
    \def\upi{\pi}
    \def\umu{\mu}
    \def\upartial{\partial}
  \fi
\fi

\title[${\rm H}_2$ Formation on Interstellar Grains]
	{${\rm H}_2$ Formation on Interstellar Grains
	in Different Physical Regimes}
\author[O. Biham et al.]
	{O.~Biham,$^1$ I.~Furman,$^1$ N.~Katz,$^1$
	V. Pirronello,$^2$ and G. Vidali$^3$\\
	$^1$Racah Institute of Physics, The Hebrew University of Jerusalem,
	Jerusalem 91904, Israel\\
	$^2$Istituto di Fisica, Universita' di Catania,
	Catania, Sicily, Italy\\
	$^3$Department of Physics, Syracuse University,
	Syracuse, NY 13244
	}

\begin{document}

\label{firstpage}

\maketitle

\begin{abstract}
An analysis of the kinetics of ${\rm H}_2$ formation on
interstellar dust grains is presented
using rate equations.
It is shown that semi-empirical expressions 
that appeared in the literature represent two different physical regimes. 
In particular, it is shown that the expression given by
Hollenbach, Werner \& Salpeter
[ApJ, 163, 165 (1971)]
applies when high flux, or high mobility,
of H  atoms on the surface of a grain,
makes it very unlikely that H atoms evaporate
before they meet each other and recombine.
The expression of Pirronello et al.\ [ApJ, 483, L131 (1997)] --
deduced on the basis of accurate measurements on realistic dust analogue --
applies to the opposite regime (low coverage and low mobility).
The implications of this analysis
for the understanding of the processes dominating in the Interstellar Medium
are discussed.
\end{abstract}


\section{Introduction}
The problem of the formation of molecular hydrogen, the most important 
species 
in the universe, is a fundamental open question in astrophysics 
\cite{duley84}.   
It was recognised long ago \cite{gould63} that 
${\rm H}_2$ cannot form in the gas phase in the Interstellar
Medium (ISM) efficiently enough to account for its abundance.
It was proposed that dust grains act as catalysts
allowing the protomolecule to quickly release 
the 4.5 eV of excess energy
(in a time comparable to the vibration
period of the highly vibrationally excited state in which it is formed).
Briefly stated, the  problem is as follows.
An H atom approaching the surface of a grain has
a probability $\xi$ (sticking coefficient) to become trapped.
The adsorbed H atom (adatom)
will spend an average time $t_{\rm H}$ (residence time)
before leaving the surface.
If during the residence time the H adatom encounters another H adatom 
(which has just landed on the surface or was already trapped in a 
deeper adsorption site),
an ${\rm H}_2$ molecule will form with a certain 
probability.
Given the fact that, until recently,
there were no experiments done in conditions relevant to the ISM,
and that little is known
of the chemical composition and morphology of dust grains,
it is not surprising that quite different models co-existed.
This is an area, as many others in astrophysics and astrochemistry, 
in which by far more theoretical papers have been written 
[Gould \& Salpeter \shortcite{gould63};
Williams \shortcite{williams68};
Hollenbach \& Salpeter \shortcite{hollenbach70,hollenbach71};
Hollenbach, Werner \& Salpeter \shortcite{hollenbach71a};
Smoluchowski \shortcite{smoluchowski81,smoluchowski83};
Aronowitz \& Chang \shortcite{aronowitz85};
Duley \& Williams \shortcite{duley86};
Buch \& Zhang \shortcite{buch91};
Sandford \& Allamandola \shortcite{sandford93}]
than experiments have been done.

A milestone has been certainly set by 
Hollenbach \& Salpeter \shortcite{hollenbach70,hollenbach71},
who treated sticking and accommodation of 
H atoms in a semiclassical way, while the mobility was treated quantum 
mechanically. They concluded that tunneling between adsorption sites, 
even at 10K, would have assured the required mobility.
Hollenbach et al.\ \shortcite{hollenbach71a} obtained 
for the steady state production rate of molecular hydrogen per unit volume
the simple expression:
\begin{equation}
\label{eq:salpeter}
	R_{\rm H_2} = {1 \over 2}
		n_{\rm H} v_{\rm H} \sigma \xi \eta n_{\rm g},
\end{equation}
where $n_{\rm H}$ and $v_{\rm H}$ are the number density and the speed
of H atoms in the gas phase respectively,
$\sigma$ the average cross-sectional area of a grain,
$n_{\rm g}$ is the number density of dust grains,
and $\eta$ is the probability that two H adatoms on the surface
meet and recombine to form ${\rm H}_2$.
Note that in the original formulation by Hollenbach et al.\
\shortcite{hollenbach71a}
$R_{\rm H_2} = (1/2) n_{\rm H} v_{\rm H} \sigma \gamma n_{\rm g}$,
where $\gamma$ is the fraction of H atoms striking the grain
that eventually form a molecule, namely, $\gamma = \xi \eta$.
Eq.\
	(\ref{eq:salpeter})
states that, for $\eta = 1$, whenever two H atoms are adsorbed on a grain,
a ${\rm H}_2$ molecule is formed.

From the experimental point of view,
besides some pioneering work in the early '60s
\cite{king63}
and '70s
\cite{schutte76},
only very recently the problem has been investigated again
\cite{pirronello97a,pirronello97b}.
Pirronello et al.\ performed their measurements
in an ultra high-vacuum (UHV) chamber 
(typical experimental pressures were in the $10^{-10}$ torr) irradiating
the sample, maintained at temperatures between 5K and 15K,
with H and D atoms  from two different triple differentially pumped lines
(D atoms were used to obtain a better signal to noise ratio).
For the very first time they used as a substrate a natural 
"olivine" (a Mg, Fe silicate) slab (mechanically polished until shiny), 
that has to be considered with good reasons a better analogue of 
interstellar dust than any model surface. A quadrupole mass spectrometer
detected the amount of HD formed on the cold substrate.
Measurements were performed both during and after irradiation with H
and D atoms. In the latter case, a Temperature Programmed Desorption
(TPD) experiment was carried out in which the temperature of the
sample is quickly ramped to over 30 K
to desorb all weakly adsorbed species.

The main results obtained by Pirronello et al.\ 
\shortcite{pirronello97a,pirronello97b} are:
(a)
In the temperature range of interest for interstellar applications
(between 10K and 15K),
the formation rates deduced from their experimental data are up to 	   
one order of magnitude lower than those calculated by 
Hollenbach \& Salpeter 
\shortcite{hollenbach70,hollenbach71}
and Hollenbach et al.\ 
\shortcite{hollenbach71a};
(b)
According to their desorption spectra, hydrogen, that is adsorbed as atomic,
becomes mobile only around 10 K, even at the high coverage regime
[see Fig.\ 2 in Pirronello et al.\ \shortcite{pirronello97b}].
Thus, at the lowest temperatures (less than about 10 K) tunneling 
alone does not provide enough mobility to H adatoms.
Instead, they find that thermal activation is required.
A possible scenario is that thermal energy is necessary to
raise H adatoms inside the adsorption well
to an energy level from which tunneling can become effective;
(c)
According to a careful analysis of the kinetics
of HD desorption spectra during TPDs,
different regimes can be recognised during ${\rm H}_2$ formation,
depending on the values of certain parameters discussed below. 
This analysis has led to the proposal of the following expression
for the steady state formation rate of ${\rm H}_2$,
\begin{equation}
\label{eq:pirronello}
	R_{\rm H_2} = {1 \over 2} (n_{\rm H} v_{\rm H} \sigma \xi t_{\rm H})^2
		n_{\rm g} \tilde{N}^{-2} \nu f(T, a, \delta E) \gamma\prime,
\end{equation}
where $\tilde{N}^2$ is the average number of hops an adatom needs to make
to encounter another adatom while performing a random walk,
and $\nu f(T, a, \delta E)$ describes the hopping rate of adatoms
due to both thermal activation and tunneling.
Here, $\nu$ is a characteristic attempt rate,
$T$ is the grain temperature,
while $a$ and $\delta E$ are width and height of the
energy barrier, respectively.
$\gamma\prime$ is the probability
that two H adatoms recombine after encountering.
In this expression the rate of ${\rm H}_2$ formation is proportional to the
square of the effective incoming H flux ($n_{\rm H} v_{\rm H} \sigma \xi$)
on the grain surface,
and is built
on a purely phenomenological basis to interpret the experimental data
on desorption kinetics.

In Eq.\
	(\ref{eq:salpeter}),
that has been used by Hollenbach et al.\ \shortcite{hollenbach71a}
and others in their chemical models,
the recombination rate is linearly proportional
to the effective incoming H flux. 
In this note it is shown that rate equations yield
Eqs.\
	(\ref{eq:salpeter})
and
	(\ref{eq:pirronello}),
as two distinct limiting cases of the H recombination rate.

\section{The Calculations}
For simplicity we will consider
the ${\rm H}_2$ production rate on a single grain;
the total production rate per unit volume can be obtained
by multiplying it by $n_{\rm g}$, the number density of dust grains.
The number of H adatoms on the surface $N_1$,
and that of ${\rm H}_2$ molecules $N_2$,
as a function of time is given by the rate equations
\begin{subeqnarray}
\label{eqs:rates}
	\dot{N}_1 & = & F - p_1 N_1 - 2 \alpha {N_1}^2
				\slabel{eq:rates1}	\\
	\dot{N}_2 & = & \alpha {N_1}^2 - p_2 N_2
				\slabel{eq:rates2}
\end{subeqnarray}
where $F$ is the rate of adsorption of H atoms on the grain,
$\alpha$ is the rate of ${\rm H}_2$ recombination (given by the product of
the diffusion coefficient and the recombination probability),
$p_1$ and $p_2$  are the desorption rates for H and ${\rm H}_2$, respectively.
The first term in Eq.\
(\ref{eq:rates1})
represents the growth of the H population due to the incoming flux;
the second term is the decrease in the H population
due to the desorption of H adatoms;
and the third term represents the rate
at which H adatoms are lost due to the recombination process.
In Eq.\
(\ref{eq:rates2})
the first term represents the rate of creation of ${\rm H}_2$ molecules.
It is related to the last term in Eq.\
(\ref{eq:rates1}) through a factor 1/2
because two H adatoms are needed to form one ${\rm H}_2$ molecule.
The second term in Eq.\
(\ref{eq:rates2})
is the desorption rate of ${\rm H}_2$ molecules
and is equal to the desired production rate
$R_{\rm H_2} = p_2 N_2$,
of ${\rm H}_2$ molecules released to the gas phase.
Note that according to this formulation
$F = n_{\rm H} v_{\rm H} \sigma \xi$, $p_1 = 1/t_{\rm H}$
and $\alpha = \tilde{N}^{-2} \nu f(T, a, \delta E) \gamma\prime$.

We will now consider the steady state conditions, where
$\dot{N}_1 = \dot{N}_2 \equiv 0$.
In this case $N_1$ can be extracted from Eq.\
(\ref{eq:rates1}),
giving rise to
$N_1 = [ - p_1 + (p_1^2 + 8\alpha F)^{1/2} ] / (4 \alpha)$,
where the unphysical negative solution is discarded.
The steady state condition and Eq.\
(\ref{eq:rates2})
imply
$R_{\rm H_2} = p_2 N_2 = \alpha N_1^2$.
By substituting the expression for $N_1$ into this equality
we find an exact formula for the ${\rm H}_2$ production rate of a single grain
\begin{equation}
\label{eq:Rexact}
	R_{\rm H_2} = 
		{p_1^2 -
		 p_1 (p_1^2 + 8 \alpha F)^{1/2} +
		 4 \alpha F
		 \over 8 \alpha} .
\end{equation}
Note that the ${\rm H}_2$ desorption rate $p_2$,
does not affect the steady state production rate $R_{\rm H_2}$.
However, it will affect the number of ${\rm H}_2$ molecules on the grain $N_2$
in the steady state.
We now evaluate expression
(\ref{eq:Rexact})
in two limiting cases.
The first case is when the adatom desorption rate
is negligible compared to their recombination rate on the surface.
This is the limit of ${p_1}^2 \ll \alpha F$.
We can neglect the first two terms
in the numerator on the right hand side of Eq.\
(\ref{eq:Rexact}),
finding
\begin{equation}
\label{eq:Rsmallp}
	R_{\rm H_2} = {1 \over 2} F ;
		\;\;\;\;\;\;\;\;\;\;\;\;\;\;\;\;
		{p_1}^2 \ll \alpha F,
\end{equation}
namely,
all H atoms that attach to the surface
recombine and desorb as ${\rm H}_2$ molecules.
In the other limit, ${p_1}^2 \gg \alpha F$, adatom desorption is important
and we expect the production rate $R_{\rm H_2}$
to be dependent on $p_1$.
Indeed, expanding the square root in Eq.\
(\ref{eq:Rexact})
according to
\(
	(1 + x)^{1/2} \simeq 1 + x/2 - x^2/8
\)
with $x = (8 \alpha F) / {p_1}^2$, we obtain:
\begin{equation}
\label{eq:Rlargep}
	R_{\rm H_2} = {\alpha \over {p_1}^2} F^2 ;
		\;\;\;\;\;\;\;\;\;\;\;\;\;\;\;\;
		{p_1}^2 \gg \alpha F .
\end{equation}
The important result of Eqs.\
(\ref{eq:Rsmallp})
and
(\ref{eq:Rlargep}),
which is the focus of this note,
is that the production rate $R_{\rm H_2}$ can be
{\em either linear or quadratic} in the adsorption rate $F$ of H atoms,
depending on the conditions stated above.
Eq.\
(\ref{eq:Rlargep})
also predicts the dependence of $R_{\rm H_2}$
on the recombination rate $\alpha$
and the adatom desorption rate $p_1$ when the latter is significant.

To make contact with Eqs.\
(\ref{eq:salpeter})
and
(\ref{eq:pirronello})
we substitute $F = n_{\rm H} v_{\rm H} \sigma \xi$,
$p_1 = 1/t_{\rm H}$
and $\alpha = \tilde{N}^{-2} \nu f(T, a, \delta E) \gamma\prime$,
and multiply the result by $n_{\rm g}$, the grain density.
Thus, Eq.\
(\ref{eq:Rsmallp})
is transformed into Eq.\
(\ref{eq:salpeter})
apart from a factor $\eta$ which is in fact equal to unity
at the conditions studied by Hollenbach et al.\
\shortcite{hollenbach71a}.
Similarly, Eq.\
(\ref{eq:Rlargep})
is identical to Eq.\
(\ref{eq:pirronello}),
apart from a factor of $1/2$ that arises
due to a different definition of $\alpha$.

\section{Discussion}
It has been shown analytically that the expressions to calculate the 
${\rm H}_2$ formation rate given in Eqs.\
(\ref{eq:salpeter})
and
(\ref{eq:pirronello})
are correct in two different regimes.
Eq.\
(\ref{eq:salpeter}),
introduced by Hollenbach et al.\ \shortcite{hollenbach71a}
almost thirty years ago,
holds when high flux, or high mobility of H adatoms,
makes it very unlikely that H adatoms evaporate before they recombine.
Specifically, it is the appropriate expression to use when
$\alpha F \gg {p_1}^2$.
The assumption of a high mobility has been experimentally proven
\cite{pirronello97a,pirronello97b}
to be inappropriate for the case in which
a more realistic grain analogue surface is used
 rather than a monocrystalline one.
Experimental data suggest that H adatoms have limited mobility and,
in the low coverage regime,
${\rm H}_2$ is not readily formed at the lowest temperature.
Such a result had been already obtained theoretically by Smoluchowski
\shortcite{smoluchowski81}.
In his calculations of ${\rm H}_2$ formation on amorphous water,
he obtained that the onset temperature for mobility and recombination
was about 18K,
a value significantly higher than 10 K measured by Pirronello et al.\ 
for olivine.
In dense, cold clouds, Smoluchowski's value would be too high 
and would make H recombination too infrequent.
The expression given in Eq.\
(\ref{eq:pirronello})
introduced by Pirronello et al.\ \shortcite{pirronello97b},
can be applied in the opposite limit or whenever the physical conditions 
maintain a low flux or low mobility of H adatoms,
with respect to their desorption rate.

\section{Implications for the interstellar medium}
The results on the rate of formation of molecular hydrogen
obtained experimentally 
\cite{pirronello97a,pirronello97b},
and also shown to hold on theoretical grounds in this communication,
can be used to speculate on how this fundamental process might proceed
during the dynamical evolution of an interstellar cloud.

As is well known,
a typical interstellar cloud crosses diffuse and dense stages
according to its energy balance with the surroundings. When a cloud
loses more energy than it gains from the environment, it contracts;
in the opposite limit, it expands.  In the evaluation of the rate of
formation of molecular hydrogen, the criteria of applicability of
Eqs.\
(\ref{eq:salpeter})
and
(\ref{eq:pirronello})
will then depend on the particular stage the
interstellar cloud is going through.
In diffuse clouds conditions favour Eq.\ (\ref{eq:pirronello}).
This is because of the low flux of H atoms (due to the low gas phase
density) and of the relatively high grain temperature (hence low H
residence time on the grain surface).
If such a diffuse cloud evolves towards a denser stage,
the conditions will probably tend to favour Eq.\ (\ref{eq:salpeter})
due to the increase of the gas phase density,
and to the decrease of the grain temperature.
In passing, we note that one must be aware
that the processes described above of ${\rm H}_2$ formation on bare refractory
grains are influenced by at least two other processes, i.e. the
decrease of H atom density in the gas phase (due to the ongoing
conversion into ${\rm H}_2$) and the accretion of an icy mantle. In the latter
case, the formation rate of ${\rm H}_2$ might be quite different than the one
on a sparsely covered silicate surface due to the different depth distributions
of adsorption sites and hence the different values of sticking and
mobility of H adatoms. It would be very useful to have laboratory
measurements of ${\rm H}_2$ formation under these conditions.

Finally, when a full dense cloud environment is developed, Eq.\
(\ref{eq:pirronello})
should apply again, because of the low density of H in the gas phase
and because of the competition between ${\rm H}_2$ and H
adatoms in occupying available adsorption sites.
Moreover, under these conditions,
the grain surface may be covered by ice
where binding energies for ${\rm H}_2$ are slightly larger than for H.
In this case, H adatoms will likely be far apart from each other
and will have to undergo a long migration before encountering each other,
which in turn favours Eq.\
(\ref{eq:pirronello}).
If we consider evolution from a dense toward a diffuse stage
the applicability of Eqs.\
(\ref{eq:salpeter})
and
(\ref{eq:pirronello})
will of course take place in the reverse order.
Quantitative studies of time dependent chemical models of interstellar clouds
have been done under both static conditions
\cite{d'hendecourt85,hasegawa93,bergin95},
and dynamically evolving conditions
\cite{charnley88a,charnley88b,nejad90,prasad91,rawlings92,shalabiea95}.
The qualitative framework we have just given
has to be confirmed by quantitatively incorporating it
into calculations describing the evolution of an interstellar cloud.

\section*{Acknowledgements}
Helpful discussions with Giulio Manic\'{o} are gratefully acknowledged.
G. Vidali would like to acknowledge support from NASA Grant NAG5-4998.
V. Pirronello would like to acknowledge support from the Italian National 
Research Council (CNR) grant CN96.00307.02.

\label{lastpage}


\begin{thebibliography}{}
  \bibitem[\protect\citename{Aronowitz \& Chang }1985]{aronowitz85}
	Aronowitz S., Chang S., 1985, ApJ, 293, 243

  \bibitem[\protect\citename{Bergin, Langer \& Goldsmith }1995]{bergin95}
	Bergin E.A., Langer W.D., Goldsmith P.D., 1995, ApJ, 441, 222

  \bibitem[\protect\citename{Buch \& Zhang }1991]{buch91}
	Buch V., Zhang Q., 1991, ApJ, 379, 647

  \bibitem[\protect\citename{Charnley et al.\ }1988a]{charnley88a}
	Charnley S.B., Dyson J.E., Hartquist T.W., Williams, D.A., 1988a,
	MNRAS, 231, 269

  \bibitem[\protect\citename{Charnley et al.\ }1988b]{charnley88b}
	Charnley S.B., Dyson J.E., Hartquist T.W., Williams, D.A., 1988b,
	MNRAS, 235, 1257

  \bibitem[\protect\citename{Duley \& Williams }1984]{duley84}
	Duley W.W., Williams D.A., 1984,
	Interstellar Chemistry, Acad. Press, London

  \bibitem[\protect\citename{Duley \& Williams }1986]{duley86}
	Duley W.W., Williams D.A., 1986, MNRAS, 223, 177

  \bibitem[\protect\citename{Gould \& Salpeter }1963]{gould63}
	Gould R.J., Salpeter E.E., 1963, ApJ, 138, 393

  \bibitem[\protect\citename{Hasegawa \& Herbst }1993]{hasegawa93}
	Hasegawa T., Herbst E., 1993, MNRAS 261, 83

  \bibitem
[\protect\citename{d'Hendecourt, Allamandola \& Greenberg}1985]{d'hendecourt85}
	d'Hendecourt L., Allamandola L.J., Greenberg J.M., 1985, A\&A, 152, 130

  \bibitem[\protect\citename{Hollenbach \& Salpeter }1970]{hollenbach70}
	Hollenbach D.J., Salpeter E.E., 1970, J. Chem. Phys., 53, 79

  \bibitem[\protect\citename{Hollenbach \& Salpeter }1971]{hollenbach71}
	Hollenbach D.J., Salpeter E.E., 1971, ApJ, 163, 155

  \bibitem
[\protect\citename{Hollenbach, Werner \& Salpeter }1971]{hollenbach71a}
	Hollenbach D.J., Werner M.W., Salpeter E.E., 1971, ApJ, 163, 165

  \bibitem[\protect\citename{King \& Wise }1963]{king63}
	King A.B., Wise H., 1963, J. Phys. Chem., 67, 1163

  \bibitem[\protect\citename{Nejad, Williams \& Charnley }1990]{nejad90}
	Nejad L.A.M., Williams D.A., Charnley S.B., 1990, MNRAS, 246, 183

  \bibitem[\protect\citename{Pirronello et al.\ }1997a]{pirronello97a}
	Pirronello V., Liu C., Shen L., Vidali G.,
	1997a, ApJ, 475, L69

  \bibitem[\protect\citename{Pirronello et al.\ }1997b]{pirronello97b}
	Pirronello V., Biham O., Liu C., Shen L., Vidali G.,
	1997b, ApJ, 483, L131

  \bibitem[\protect\citename{Prasad, Heere \& Tarafdar }1991]{prasad91}
	Prasad S.S, Heere K.R., Tarafdar S.P. 1991, ApJ, 373, 123

  \bibitem[\protect\citename{Rawlings et al.\ }1992]{rawlings92}
	Rawlings J.M.C., Hartquist T.W., Menten K.M., Williams D.A., 1992,
	MNRAS, 255, 471

  \bibitem[\protect\citename{Sandford \& Allamandola }1993]{sandford93}
	Sandford S.A., Allamandola L.J., 1993, ApJ, 409, L65

  \bibitem[\protect\citename{Schutte et al.\ }1976]{schutte76}
	Schutte A., Bassi D., Tommasini F., Turelli F.,
	Scoles G., Herman L.J.F.,
	1976, J. Chem. Phys., 64, 4135

  \bibitem[\protect\citename{Shalabiea \& Greenberg }1995]{shalabiea95}
	Shalabiea O.M., Greenberg J.M., 1995, A\&A, 303, 233

  \bibitem[\protect\citename{Smoluchowski }1981]{smoluchowski81}
	Smoluchowski R., 1981, Ap\&SS, 75, 353

  \bibitem[\protect\citename{Smoluchowski }1983]{smoluchowski83}
	Smoluchowski R., 1983, J. Phys. Chem., 87, 4229

  \bibitem[\protect\citename{Williams }1968]{williams68}
	Williams D.A., 1968, ApJ, 151, 935

\end{thebibliography}
\end{document}